\PassOptionsToPackage{table,xcdraw}{xcolor}
\documentclass[manuscript]{acmart}
\usepackage{tabularx}
\usepackage{multirow}
\usepackage{subcaption}

\setcopyright{acmcopyright}
\copyrightyear{2025}
\acmYear{2025}
\acmDOI{XXXXXXX.XXXXXXX}
\acmBooktitle{CUI '25} 
\acmISBN{978-1-4503-XXXX-X/18/06}

\begin{document}

\title{Personas Evolved: Designing Ethical LLM-Based Conversational Agent Personalities}

\author{Smit Desai}
\orcid{0000-0001-6983-8838}
\affiliation{%
  \institution{Northeastern University}
  \country{USA}}
\email{sm.desai@northeastern.edu}

\author{Mateusz Dubiel}
\orcid{0000-0001-8250-3370}
\affiliation{%
  \institution{University of Luxembourg}
  \country{Luxembourg}}
\email{mateusz.dubiel@uni.lu}

\author{Nima Zargham}
\email{zargham@uni-bremen.de}
\orcid{0000-0003-4116-0601}
\affiliation{
  \institution{Digital Media Lab, University of Bremen}
  \country{Germany}
}

\author{Thomas Mildner}
\email{mildner@uni-bremen.de}
\orcid{0000-0002-1712-0741}
\affiliation{
    \institution{Digital Media Lab, University of Bremen}
    \country{Germany}
}

\author{Laura Spillner}
\email{laura.spillner@uni-bremen.de}
\orcid{0000-0001-8490-8961}
\affiliation{
    \institution{Digital Media Lab, University of Bremen}
    \country{Germany}
}

\renewcommand{\shortauthors}{Desai et al.}
\begin{abstract}
The emergence of Large Language Models (LLMs) has revolutionized Conversational User Interfaces (CUIs), enabling more dynamic, context-aware, and human-like interactions across diverse domains, from social sciences to healthcare. However, the rapid adoption of LLM-based personas raises critical ethical and practical concerns, including bias, manipulation, and unforeseen social consequences. Unlike traditional CUIs, where personas are carefully designed with clear intent, LLM-based personas generate responses dynamically from vast datasets, making their behavior less predictable and harder to govern. This workshop aims to bridge the gap between CUI and broader AI communities by fostering a cross-disciplinary dialogue on the responsible design and evaluation of LLM-based personas. Bringing together researchers, designers, and practitioners, we will explore best practices, develop ethical guidelines, and promote frameworks that ensure transparency, inclusivity, and user-centered interactions. By addressing these challenges collaboratively, we seek to shape the future of LLM-driven CUIs in ways that align with societal values and expectations.
\end{abstract}

\maketitle

\section{Theme and Goals}
Designing unique personas and identities for agents—which serve as the front-end or the interaction layer—has long been an integral part of conversational user interfaces (CUI). As intermediaries between users and conversational agents, CUI designers have crafted personas—distinct from traditional HCI personas—to fulfill specific roles, such as doctors providing health information or physical trainers offering personalized guidance \cite{desai2023metaphors, Desai_Chin_2023, Motalebi_Cho_Sundar_Abdullah_2019, Desai_Hu_Lundy_Chin_2023, Chin_Desai_2021}. These personas have played a key role in creating tailored, user-centered experiences. However, the emergence of Large Language Models (LLMs) has introduced a significant shift in this paradigm.

LLMs enable natural, coherent dialogues and extensive knowledge of the world. This capability allows LLMs to simulate human behavior in text-based contexts, giving rise to LLM Agents \cite{Park_O’Brien_Cai_Morris_Liang_Bernstein_2023}. These agents provide a technical foundation for designing advanced CUIs that are more dynamic and context-aware than previous systems. Additionally, LLMs have introduced the possibility of role-playing as humans in complex, text-based scenarios. 
These models have revolutionized the creation of agent personas through advanced prompting techniques, as evidenced by numerous contributions from the machine-learning community. For example, Park et al. \cite{Park_O’Brien_Cai_Morris_Liang_Bernstein_2023} developed a virtual village populated by generative agents, each assigned distinct roles—artists create paintings, while authors compose literary works. These agents autonomously interact within the environment, simulating complex social behaviors by incorporating detailed persona inputs—such as life context, preferences, cognition, and emotions—to predict human behaviors and cognitive patterns to create dynamic, evolving digital ecosystems with unprecedented realism and flexibility \cite{Park_O’Brien_Cai_Morris_Liang_Bernstein_2023, Suzuki_Arita_2024, Serapio-García_Safdari_Crepy_Sun_Fitz_Romero_Abdulhai_Faust_Matarić_2023, Jiang_Zhang_Cao_Breazeal_Roy_Kabbara_2024}. 
This potential is increasingly being realized in virtual applications and events. For instance, Shoa et al. \cite{Shoa_Oliva_Slater_Friedman_2023} documented a hybrid panel session during a physical symposium, where an LLM-based virtual agent embodied an Albert Einstein persona and engaged in a live discussion alongside three international experts. Such examples illustrate how LLMs can seamlessly integrate into dynamic, real-world environments, offering new possibilities for interactive and immersive experiences.

However, alongside these advancements, the integration of LLM-based personas raises significant ethical and practical concerns. The ability to generate complex, human-like personas with minimal effort introduces risks related to user manipulation, misinformation, and unintended emotional attachments. For instance, character.ai\footnote{https://character.ai/}, a California-based company, allows users to interact with pre-designed personas such as Best Friend, Boss or CEO, Moo Deng, and specific Adult Film Stars or create their own by simply filling in four prompt fields: ``Character Name,''``Description,'' ``Tagline,'' and ``Greeting.'' While this flexibility enables users to explore AI-generated personas with ease, it also grants unprecedented power over digital interactions. Tragically, this lack of safeguards has led to serious real-world consequences. In one case, a U.S. teenager engaged with a persona modeled after ``Daenerys'' from \textit{Game of Thrones}, who reportedly instructed the teenager to ``come home to her,'' resulting in fatal consequences\footnote{https://www.nytimes.com/2024/10/23/technology/characterai-lawsuit-teen-suicide.html}.

The challenge of designing ethical agent personas is not new. The CUI community has extensive experience in crafting personas for systems acting as doctors, therapists, storytellers, and coaches \cite{desai2023metaphors}. Over the years, research in this space has explored ethical concerns such as manipulation, user dependency, and the unintended consequences of anthropomorphization in past CUI conferences \cite{Desai_Twidale_2022, Simpson_Crone_2022a, Pradhan_Lazar_2021a}. However, the emergence of LLM-driven conversational models represents a paradigm shift that requires revisiting these concerns. Unlike traditional intent-based architectures reliant on Natural Language Understanding (NLU) models, LLM-based systems generate responses dynamically, introducing new challenges in design, evaluation, and deployment. While LLMs offer unparalleled flexibility, creativity, and scalability, they also complicate existing best practices, making it necessary to adopt hybrid approaches that leverage LLMs' capabilities while mitigating their risks.

Additionally, there remains a lack of consistency in how researchers, designers, and developers describe these systems. Terms such as persona, agent, and character are often used interchangeably, leading to ambiguity in discussions and implementations \cite{Chen_Yao_Ye_Wang_Li_2024}. This inconsistency highlights the need for a shared vocabulary that can facilitate more precise communication across disciplines and support the development of clear ethical frameworks.

The goal of this workshop is to initiate a dialogue between the CUI and broader AI communities to critically examine the design and ethical implications of LLM-based personas. Specifically, this workshop will facilitate discussions on the current state of literature and best practices for designing personas, drawing from research in conversational user interfaces, machine learning, and related fields. Additionally, it aims to establish a shared vocabulary by clarifying overlapping terms such as \textit{persona}, \textit{agent}, and \textit{character}, improving cross-disciplinary communication. Furthermore, the workshop will explore the ethical dimensions of persona design, identifying key risks and challenges while laying the groundwork for future research on guidelines and best practices. By bringing together researchers, designers, and practitioners from human-computer interaction, artificial intelligence, social robotics, and other relevant fields, this workshop seeks to foster collaboration, outline research opportunities, and develop a common foundation for ethical and user-centered LLM-based personas. 



\section{Organisers}
\noindent \textbf{Smit Desai} is an Assistant Professor at Northeastern University's College of Art, Media and Design. He studies user mental models of conversational agents using innovative methods like metaphor analysis. He applies this insight and his expertise in conversation design to develop voice-based AIs for social roles such as instructors and coaches. Additionally, he has served the HCI community in various roles, such as co-chair of Provocation Papers at CUI 2024 and co-organizer for workshops at CHI 2024, MUM 2024, and CUI 2024. 

\noindent \textbf{Mateusz Dubiel} is a Research Associate in the Department of Computer Science at the University of Luxembourg, where he works on the development and evaluation of conversational agents. Specifically, his current research focuses on assessment of cognitive and usability implications of interfaces that feature speech and exploration of their potential to inspire positive behavioural change in users. He served as Short Papers Chair for CUI 2022 and was one of General Chairs for CUI 2024.

\noindent \textbf{Nima Zargham} is a postdoctoral researcher in the Digital Media Lab at the University of Bremen. His research focuses on human-centered approaches for designing speech-based systems that elicit desirable user experiences. Nima has previously organized CUI-related workshops at notable conferences such as ACM/IEEE HRI 2023, ACM CUI 2023, ACM CUI 2024, and ACM CHI 24. Additionally, he served as a local chair at the ACM CHI-PLAY 2022 conference. His research efforts have resulted in publications featured in prestigious HCI venues, including ACM's CHI, CUI, and CHI-PLAY. 

\noindent \textbf{Thomas Mildner} is a postdoctoral researcher at the Digital Media Lab at the University of Bremen. His research focuses on ethical and responsible design and online well-being, with studies exploring so-called dark patterns in social media as well as conversational technologies. To this end, Thomas collaborated to develop the CUI expectation cycle to align system capabilities with user expectations~\cite{mildner_listening_2024} and an ontology for dark patterns~\cite{gray_ontology_2024}. His research has been published in venues including CHI, DIS, and CUI.

\noindent \textbf{Laura Spillner} is a PhD student at the Digital Media Lab at the University of Bremen. Her research focuses on the development of explainable AI and hybrid AI in the field of language understanding and the interaction between humans and AI in natural language.

\section{Plans for the Workshop}

\subsection{Pre-workshop Plans}
To maximize awareness and participation, we will promote the workshop through multiple channels, including the organizers’ professional networks, social media platforms, and relevant academic and industry mailing lists. A dedicated workshop website (TBA) will serve as a central hub, providing details such as submission guidelines, key deadlines, and logistical information for attendees. Prospective participants will be invited to submit either a position paper (2-3 pages) or a statement of interest. Priority will be given to those submitting position papers, as they will help ground the workshop discussions in concrete ideas and research. To ensure a productive setting, participation will be limited to 20 individuals. The selection process will prioritize diversity, aiming to assemble a group that reflects a wide range of backgrounds, expertise, and perspectives relevant to the workshop’s themes. By carefully curating the participant list, we aim to create an environment conducive to meaningful dialogue, collaboration, and knowledge exchange.

\subsection{Tentative Schedule and Workshop Activites}
The tentative schedule and workshop activities are presented in \autoref{tab:activities}. Since the purpose of our workshop is to synthesize insights from different communities and translate them into practical persona design and evaluation recommendations, we will ensure that ample time is dedicated to prototyping activities, discussions, and interactions between the participants. We further aim to enable networking and promote collaborations on this topic, encouraging participants to build connections and continue exploring the themes discussed beyond the scope of this workshop.

\begin{table}[h!]
\caption{Tentative Workshop Plan}
\begin{tabular}{lll}
\toprule
\multicolumn{1}{c}{\begin{tabular}[c]{@{}c@{}}\textbf{Time  Slot (duration)} \end{tabular}} & \multicolumn{1}{c}{\begin{tabular}[c]{@{}c@{}}\textbf{Activity}\end{tabular}} &  \\ \midrule
09:00-09:15~(15mins) & Welcome and Introductions &  \\
09:15-9:45~(60mins) & Insights and Examples  &  \\
09:45-10:30~(15mins) & Prototyping Personas  &  \\
10:30-10:45~(15mins) & Coffee Break  &  \\
10:45-11:30~(45mins) & Persona Implementing and Evaluation &  \\
11:30-12:00~(30mins) & Keynote by Dr. Marta Ziosi &  \\
12:00-12:30~(30mins) & Closing Remarks and Future Plans  &  \\
\bottomrule
\end{tabular}
\label{tab:activities}
\end{table}

At the start of the workshop, the organizers will present examples of LLM-based personas, followed by a discussion on their design and key characteristics. This will set the stage for a hands-on group prototyping session, where participants will collaborate to design their own LLM-based persona for a specific application.
To support this process, each group will be guided by at least one workshop organizer. Participants will have 30 minutes to develop their personas and prepare brief presentations explaining their design choices. They will then deploy their personas on an existing LLM, such as ChatGPT-o1 or DeepSeek r1.
Following deployment, each group’s persona will be tested and evaluated by another group. This will lead to a discussion on the experience of interacting with the personas and insights into their design process. Finally, key takeaways will be summarized and presented. The format and duration of the reporting and discussion will be adjusted based on the number of attendees (maximum 20 participants).

\subsubsection{Keynote by Dr. Marta Ziosi}
To inspire discussions, we are delighted that Dr. Marta Ziosi has agreed to give a keynote at this event.
Marta is an AI policy expert who worked for institutions such as DG CNECT at the European Commission, the Berkman Klein Centre for Internet \& Society at Harvard University, The Montreal International Center of Expertise in Artificial Intelligence (CEIMIA) and The Future Society.
Previously, Marta was a Ph.D. student and researcher on Algorithmic Bias and AI Policy at the Oxford Internet Institute. She is also the founder of AI for People, a non-profit organisation whose mission is to put technology at the service of people. Currently, Marta serves as vice-chair to co-develop the General-Purpose AI Code of Practice of the European Commission, specifically risk identification and assessment, including evaluations.


\subsubsection{Commitment to Diversity and Inclusivity}
Our goal is to create an environment where everyone feels valued and included. We are dedicated to ensuring broad participation by reaching out to individuals from diverse backgrounds, particularly those underrepresented in HCI and AI. Accessibility is a priority—we will work with the conference organizers to make the event as inclusive as possible. To maintain a respectful and productive atmosphere, we will establish clear guidelines for conduct and provide multiple ways for participants to voice concerns or feedback. We also want to encourage meaningful connections. Through structured breakout sessions and mentorship opportunities, we will create spaces for cross-disciplinary collaboration and learning.

\subsection{Post-workshop plans}
The workshop outcomes will be compiled into a comprehensive report, which will be published on the workshop’s website to ensure broad accessibility. To extend its reach, the organizers will leverage social media platforms such as Mastodon and X (formerly Twitter) for dissemination, engaging a wider audience within the research community and beyond. Additionally, to foster continued dialogue and collaboration, we will establish a dedicated mailing list or a Discord group, depending on participant preferences. This space will serve as a hub for sharing relevant updates, discussing emerging challenges, and facilitating ongoing interdisciplinary engagement.

\begin{acks}
The authors are grateful to Dr. Bingsheng Yao for supporting the writing of an earlier version of this proposal. 
\end{acks}

\bibliographystyle{ACM-Reference-Format}
\bibliography{references}

\end{document}